\documentclass[preprintnumbers, prd, twocolumn, showpacs, floatfix, preprintnumbers, letterpaper, superscriptaddress,nofootinbib]{revtex4}
\usepackage{graphicx,epsfig}
\usepackage{amsmath}
\usepackage{amssymb}
\usepackage{bm}
\usepackage{amsfonts}
\usepackage{epstopdf}
\usepackage{subfigure}
\usepackage{multirow}

\usepackage{relsize}
\usepackage{relsize}

\begin{document}

\title{Action functional of the Cardassian universe}

\author{Xiang-hua Zhai}
\email{zhaixh@shnu.edu.cn}
\author{Rui-hui Lin}
\author{Chao-jun Feng}
\email{fengcj@shnu.edu.cn}
\author{Xin-zhou Li}
\email{kychz@shnu.edu.cn}

\affiliation{Shanghai United Center for Astrophysics(SUCA), Shanghai
Normal University, 100 Guilin Road, Shanghai 200234, China
}%

\begin{abstract}
It is known that the Cardassian universe is successful in describing the accelerated expansion of the universe, but its dynamical equations are hard to get from the action principle. In this paper, we establish the connection between the Cardassian universe and $f(T, \mathcal{T})$ gravity, where $T$ is the torsion scalar and $\mathcal{T}$ is the trace of the matter energy-momentum tensor. For dust matter, we find that the modified Friedmann equations from $f(T, \mathcal{T})$ gravity can correspond to those of Cardassian models, and thus, a possible origin of Cardassian universe is given. We obtain the original Cardassian model, the modified polytropic Cardassian model, and the exponential Cardassian model from the Lagrangians of $f(T,\mathcal{T})$ theory. Furthermore, by adding an additional term to the corresponding Lagrangians, we give three generalized Cardassian models from $f(T,\mathcal{T})$ theory. Using the observation data of type Ia supernovae, cosmic microwave background radiation, and baryon acoustic oscillations, we get the fitting results of the cosmological parameters and give constraints of model parameters for all of these models.

\end{abstract}

\maketitle

\section{Introduction}\label{Introduction}

The Cardassian universe \cite{Freese,Wang,Freese2,Gondolo} has been known to describe the accelerating expansion of the universe with remarkable agreement with observations, whereas it lacked a solid theoretical foundation until now. In Cardassian models, the Friedmann equation is modified by the introduction of an additional nonlinear term of energy density while without the introduction of a cosmological constant or any dynamic dark energy component. In these models, the universe can be flat and yet consist of only matter and radiation and still be compatible with observations. Matter can be sufficient to provide a flat geometry. The possible origin for Cardassian models is from the consideration of braneworld scenarios, where our observable universe is a three dimensional membrane embedded in extra dimensions\cite{Chung}. The modified Friedmann equation may result from the existence of extra dimensions, but it is difficult to find a simple higher dimensional theory, i. e., a higher-dimensional momentum tensor that produces the Cardassian cosmology\cite{Freese3}. Inspired by the study on the correspondence between thermodynamic behavior and gravitational equations, a couple of us have studied the thermodynamic origin of the Cardassian universe\cite{Feng}. However, it is still hard to get the dynamic equations of this model from the action principle.

 To explain the accelerated expansion of the universe, besides adding Cardassian terms or unknown fields such as quintessence \cite{Peebles,Li} and phantom \cite{Caldwell,Li2}, there is another kind of theory known as modified gravity, which uses an alternative gravity theory instead of Einstein's theory, such as $f(R)$ theory \cite{Nojiri,Du}, MOND cosmology \cite{Zhang}, Poincar\'{e} gauge theory \cite{Li3,Ao,Ao2}, and de Sitter gauge theory \cite{Ao3}. On the other hand, Einstein constructed the "Teleparallel Equivalence of General Relativity" (TEGR) which is equivalent to the general relativity (GR) from the Einstein-Hilbert action\cite{Einstein,Unzicker,Moller,Pellegrini,Aldrovandi}. In TEGR, the curvatureless Weitzenb\"{o}ck connection takes the place of the torsionless Levi-Civita one, and the vierbein is used as the fundamental
field instead of the metric. In the Lagrangian of TEGR, the torsion scalar $T$, from contractions of the torsion tensor, takes the place of the curvature scalar $R$. The simplest approach in TEGR to modify gravity is $f(T)$ theory \cite{Ferraro,Linder}, whose important advantage is that the field equations are second order and not fourth order as in $f(R)$ theory. Recently, we established two concrete $f(T)$ models that do not change the successful aspects of the Lambda cold dark matter scenario under the error band of fitting values, as describing the evolution history of the universe including the radiation-dominated era, the matter-dominated era, and the present accelerating expansion \cite{Zhai}. We also considered the spherical collapse and virialization in $f(T)$ gravities\cite{Zhai2}. Furthermore, extensions of $f(T,\mathcal{T})$ theory\cite{Harko0} where $\mathcal{T}$ is the trace of the matter energy-momentum tensor $\mathcal{T}_{\mu\nu}$ were constructed, whose cosmological implications are rich and varied.

Recently, it was shown\cite{Katirci} that modified gravity models may lead to
a Cardassian-like expansion.
In this paper, we try to find the relation between Cardassian models and $f(T,\mathcal{T})$ theory. Under the reconsidered scheme of $f(T,\mathcal{T})$ theory, we obtain  the original Cardassian model, the modified polytropic Cardassian model, and the exponential Cardassian model through the action principle and thus give a possible origin of the Cardassian universe. Furthermore, by adding an additional term to the corresponding Lagrangians, we give three generalized Cardassian models from $f(T,\mathcal{T})$ theory. Using the observation data of type Ia supernovae(SNeIa), cosmic microwave background radiation(CMB), and baryon acoustic oscillations(BAO), we get the fitting results of the cosmological parameters and give constraints of model parameters for all of these models.

The paper is organized as follows: in Sec. II, with the discussion of the self-consistent form of the Lagrangian of barotropic perfect fluid, we give a new derivation of $f(T,\mathcal{T})$ theory. In Sec. III, the actions of the three Cardassian models from $f(T,\mathcal{T})$ theory are given explicitly and generalized Cardassian models from $f(T,\mathcal{T})$ theory are further given. We also examine the observational constraints of each model in this section.  Finally, Sec. IV is devoted to the conclusion and discussion.
We use the signature convention (+,-,-,-) in this paper.

\section{$f(T,\mathcal{T})$ theory with barotropic perfect fluid}
\subsection{The Lagrangian of barotropic perfect fluid}
There exist two types of Lagrangian $\mathcal{L}_m$ for the perfect fluid in modified gravity theories, so we have to define one or the other of these two. Harko has pointed out that $\mathcal{L}_m=\epsilon(\rho)$ is a more reasonable choice \cite{Harko6} in modified gravity theories, where $\epsilon(\rho)$ is the total energy density of the fluid and $\rho$ is the rest mass density.

In the work of Brown\cite{Brown}, it is shown that the on shell perfect-fluid Lagrangian in GR can be $\mathcal{L}_m=\rho$ or $\mathcal{L}_m=-p$, where $\rho$ is the rest mass density and $p$ is the pressure. Both Lagrangians lead to the same perfect fluid stress-energy tensor concordant with the laws of thermodynamics and hence, the same equations of motion. In past years, some authors have adopted some specific form of $\mathcal{L}_m=-p$ from the work of Brown for their alternative theories of gravity \cite{Harko3,Katirci,Bertolami1,Sotiriou2,Faraoni,Farajollahi,Harko4}. However, according to Refs. \cite{Harko6,Bertolami2}, we have to reconsider how to take the form of $\mathcal{L}_m$ for the perfect fluid in modified gravity theories, including $f(T,\mathcal{T})$ theory.

The usual form of the stress tensor of a barotropic perfect fluid is
\begin{equation}\label{energytesnsor}
\mathcal{T}^{\mu \nu}=-[\epsilon(\rho)+p(\rho)]u^{\mu}u^{\nu}+p(\rho)g^{\mu \nu}
\end{equation}
where $\epsilon(\rho)$ and $p(\rho)$ are the total energy density and the pressure of the fluid, respectively, which both depend on the rest mass density $\rho$. On the other hand, if the Lagrangian of a barotropic perfect fluid $\mathcal{L}_m$ does not depend on the derivatives of the metric, the usual definition of the stress-energy tensor $\mathcal{T}^{\mu \nu}$
\begin{equation}\label{energytensor2}
\mathcal{T}^{\mu \nu}=-\mathcal{L}_m g_{\mu \nu}+ 2\frac{\partial\mathcal{L}_m}{\partial g^{\mu \nu}}
\end{equation}
where $\mathcal{L}_m$ can be assumed to depend on $\rho$ only. Considering the conservation of the matter current $\nabla_{\sigma} (\rho u^{\sigma})=0$, one can prove that\cite{Harko6,Fock}
\begin{equation}\label{densityderiation}
\delta \rho=\frac 1 2 \rho(g_{\mu \nu}-u_{\mu}u_{\nu})\delta g^{\mu \nu}
\end{equation}
where the four velocity of the fluid $u^{\alpha}$ satisfies the conditions $u^{\alpha}u_{\alpha}=1$. Substituting these results into Eq.(\ref{energytensor2}), one can obtain\cite{Harko6,Minazzoli}
\begin{equation}\label{energytensor3}
\mathcal{T}^{\mu \nu}=-\rho\frac{d\mathcal{L}_m}{d\rho}u^{\mu}u^{\nu}-\left (\mathcal{L}_m-\rho\frac{d\mathcal{L}_m}{d\rho}\right )g^{\mu \nu}.
\end{equation}
From a comparison of Eq.(\ref{energytesnsor}) and Eq.(\ref{energytensor3}), we have
\begin{equation}
\mathcal{L}_m=\epsilon(\rho)=\rho [c^2+\int p(\rho)/\rho^2d\rho].
\label{lm}
\end{equation}
and
\begin{equation}\label{derivativeenergytodensity}
\frac{d\epsilon(\rho)}{d\rho}=\frac{\epsilon(\rho)+p(\rho)}{\rho},
\end{equation}
where $c$ is the speed of light, and the unit $c=1$ is taken hereinafter. In other words, $\mathcal{L}_m=\epsilon(\rho)$ is a direct and reasonable generalization from $\mathcal{L}_m=\rho$ in GR to $f(T,\mathcal{T})$ theory, because Brown's argument becomes invalid in modified gravity theories. When compared with it, $\mathcal{L}_m=-p$ is only a direct employment from GR.

Furthermore, we can verify the conservation of the total energy . Actually, one can easily obtain the divergence of the energy density current
\begin{equation}
	\nabla_\sigma\left( \epsilon u^\sigma \right)=\left( 1+\int\frac p{\rho^2}d\rho+\frac p{\rho} \right)\nabla_\sigma\left( \rho u^\sigma \right)-p\nabla_\sigma u^\sigma.
	\label{ecurrent}
\end{equation}
Under the conservation of matter current $\nabla_\sigma\left( \rho u^\sigma \right)=0$, Eq.(\ref{ecurrent}) is the conservation of the total energy. For example, under the Friedmann-Walker-Robertson (FRW) metric it becomes
\begin{equation}
	\dot\epsilon+3H\left( \epsilon+p \right)=0,
	\label{econserv}
\end{equation}
which is the usual form of energy conservation in cosmology.

\subsection{The field equations in $f(T,\mathcal{T})$ Theory}

We can find a set of smooth basis vector fields $\hat{e}_{(\mu)}$ in different patches of the manifold $\mathcal{M}$ and make sure things are well-behaved on the overlaps as usual, where Greek indices run over the coordinates of spacetime. The set of vectors $\textbf{e}_A$, comprising an orthonormal basis, is known as a tetrad or vierbein, where Latin indices run over the tangent space $T_p$ at each point $p$ in $\mathcal{M}$. A natural basis of $T_p$ is given by $\hat{e}_{(A)}=\partial / \partial x^A$. Any vector can be expressed as  linear combinations of the basis vector, so we have

\begin{equation}
\hat{e}_{(A)}=e_A ^{\hspace{0.2cm}\mu}\hat{e}_{(\mu)}
\end{equation}
where the components $e_A^{\hspace{0.2cm}\mu}$ form a $4\times 4$ invertible matrix. We will also refer to $e_A^{\hspace{0.2cm}\mu}$ as the vierbein in  accordance with the usual practice of blurring the distinction between objects and their components. The vectors $\hat{e}_{(\mu)}$ in terms of  $\hat{e}_{(A)}$ are
\begin{equation}
\hat{e}_{(\mu)}=e^A _{\hspace{0.2cm}\mu}\hat{e}_{(A)}
\end{equation}
where the inverse vierbeins $e^A _{\hspace{0.2cm}\mu}$ satisfy
\begin{equation}
e^A_{\hspace{0.2cm}\mu}e_B^{\hspace{0.2cm\mu}}=\delta_B^A,e_A^{\hspace{0.2cm}\mu}e^A_{\hspace{0.2cm}\nu}=\delta_\nu^\mu.
\end{equation}
Therefore, the metric is obtained from $e^A_{\hspace{0.2cm}\mu}$
\begin{equation}
g_{\mu\nu}=\eta_{AB}e^A_{\hspace{0.2cm}\mu}e^B_{\hspace{0.2cm}\nu},
\end{equation}
or equivalently
\begin{equation}\eta_{AB}=g_{\mu\nu}e_A^{\hspace{0.2cm}\mu}e_B^{\hspace{0.2cm}\nu},
\end{equation}
and the root of the metric determinant is given by $|e|=\sqrt{-g}=\det(e^A_{\hspace{0.2cm}\mu})$.

In TEGR, one uses the standard Weitzenb\"{o}ck's connection defined as
\begin{equation}
\Gamma^\alpha_{\mu\nu}=e_A^{\hspace{0.2cm}\alpha}\partial_\nu e^A_{\hspace{0.2cm}\mu}=-e^A_{\hspace{0.2cm}\mu}\partial_\nu e_A^{\hspace{0.2cm}\alpha},
\end{equation}
and the covariant derivative $\mathrm{D}_\mu$ satisfies the equation
\begin{equation}
\mathrm{D}_\mu e^A_{\hspace{0.2cm}\nu}=\partial_\mu e^A_{\hspace{0.2cm}\nu}-\Gamma^\alpha_{\nu\mu}e^A_{\hspace{0.2cm}\alpha}=0.
\end{equation}
Then the components of the torsion and contorsion tensors are given by
\begin{eqnarray}
T^\alpha_{\hspace{0.2cm}\mu\nu}&=&\Gamma^\alpha_{\nu\mu}-\Gamma^\alpha_{\mu\nu}=e_A^{\hspace{0.2cm}\alpha}(\partial _\mu e^A_{\hspace{0.2cm}\nu}-\partial_\nu e^A_{\hspace{0.2cm}\mu}),\\
K^{\mu\nu}_{\hspace{0.3cm}\alpha}&=&-\frac 1 2 (T^{\mu\nu}_{\hspace{0.3cm}\alpha}-T^{\nu\mu}_{\hspace{0.3cm}\alpha}-T_\alpha^{\hspace{0.2cm}\mu\nu}).
\end{eqnarray}
By introducing another tensor
\begin{equation}
S_\alpha^{\hspace{0.2cm}\mu\nu}=\frac 1 2(K^{\mu\nu}_{\hspace{0.3cm}\alpha}+\delta ^\mu _\alpha T^{\beta\nu}_{\hspace{0.3cm}\beta}-\delta ^\nu_\alpha T^{\beta\mu}_{\hspace{0.3cm}\beta}),
\end{equation}
we can define the torsion scalar as
\begin{equation}
T\equiv T^\alpha_{\hspace{0.2cm}\mu\nu}S_\alpha^{\hspace{0.2cm}\mu\nu}.
\end{equation}

The action for $f(T,\mathcal{T})$ gravity takes the following form \cite{Harko0}
\begin{equation}
S=\frac 1{16\pi G}\int e f(T,\mathcal{T})d^4 x+\int e \mathcal{L}_m d^4 x,
\end{equation}
where $f(T, \mathcal{T})$ is an arbitrary function of the torsion scalar $T$ and the trace $\mathcal{T}$ of the matter stress-energy tensor. On the variation with respect to the vierbein that leads to the field equations, a question that should be noted is how to deal with the variation of the trace of the energy-momentum tensor $\delta \mathcal{T}$. This question has been met in theories with $\mathcal{T}$ included in the action, including $f(R, \mathcal{T})$ theory \cite{Harko3,Katirci} and $f(T, \mathcal{T})$ theory \cite{Harko0}. With the discussion in the last subsection, we can reexamine this question now.

From Eq. (\ref{densityderiation}) and Eq.(\ref{derivativeenergytodensity}), the variation of $\epsilon$ is
\begin{equation}\label{energyvariation}
\delta\epsilon=-\frac 1 2(\epsilon+p)(g^{\alpha\beta}-u^{\alpha}u^{\beta})\delta g_{\alpha\beta}.
\end{equation}
Using (\ref{energytesnsor}), (\ref{derivativeenergytodensity}), and (\ref{energyvariation}), one can express the variation of $\mathcal{T}$ as
\begin{eqnarray}
\delta \mathcal{T}&=&\delta(3p-\epsilon)\nonumber\\
&=&(3\frac{d p}{d\rho}\frac{\rho}{\epsilon+p}-1)\delta\epsilon\nonumber\\
&=&(1-3\frac{d p}{d\rho}\frac{\rho}{\epsilon+p})(\mathcal{T}^{\alpha}_{\beta}+\epsilon\delta^{\alpha}_{\beta})e_A^{\beta}\delta e_{\alpha}^A.
\end{eqnarray}
The field equations then read as
\begin{eqnarray}\label{fieldequation}
& &f e_A^{\hspace{0.2cm}\alpha}+\frac 4 e f_T \partial_{\beta}(e S_{\sigma}^{\hspace{0.2cm}\alpha\beta}e_A ^{\hspace{0.2cm}\sigma})+4S_{\sigma}^{\hspace{0.2cm}\alpha\beta}e_A^{\hspace{0.2cm}\sigma}\partial_{\beta}f_T\nonumber\\
&+&4f_TS_{\rho}^{\hspace{0.2cm}\alpha\sigma}T^{\rho}_{\hspace{0.2cm}\sigma\beta}e_A^{\hspace{0.2cm}\beta}+f_\mathcal{T} (1-3\frac{d p}{d\rho}\frac{\rho}{\epsilon+p})\epsilon e_A^{\hspace{0.2cm}\alpha}\nonumber\\
&=&\left(f_\mathcal{T}(3\frac{d p}{d\rho}\frac{\rho}{\epsilon+p}-1)+16\pi G\right)\mathcal{T}^{\alpha}_{\hspace{0.2cm}\beta} e_A^{\hspace{0.2cm}\beta}
\end{eqnarray}
where $f_T$ and $f_\mathcal{T}$ denote derivatives with respect to torsion scalar $T$ and the trace of $T^{\mu\nu}$, respectively.

In contrast to $f(T,\mathcal{T})$ theory in previous papers\cite{Harko0,Gomez,Farrugia,Pace}, this is the new derivation of $f(T, \mathcal{T})$ theory with $\delta \mathcal{T}$ reconsidered, since we have taken $\mathcal{L}_m=\epsilon(\rho)$ but not $\mathcal{L}_m=-p$. The crucial difference lies in the different choice of the matter Lagrangian $\mathcal{L}_m$. The derivation of the field equations in the references mentioned above depends on the assumption that $\mathcal{L}_m=-p$. And the same assumption is used in works on $f(R,\mathcal{T})$ gravity (see \cite{Harko3}). However, from the discussion in Sec. IIA and also in Refs. \cite{Harko6,Minazzoli}, $\mathcal{L}_m=\epsilon(\rho)$ would be a more reasonable choice. This is what leads to the difference between the field equations (\ref{fieldequation}) that we got and the ones in the literature.

Since $f(T)$ theories are known to violate local Lorentz invariance\cite{Barrow,Sotiriou}, particular choices of tetrad are important to get viable models in $f(T)$ cosmology, as has been noticed in Ref. \cite{Tamanini}. For a flat FRW metric in Cartesian coordinates,
\begin{equation}
ds^2=dt^2-a(t)^2(dx^i)^2
\label{FRWmetric}
\end{equation}
where $a(t)$ is the scale factor, the diagonal tetrad $e^A_{\hspace{0.2cm}\mu}=\mathrm{diag}(1,a,a,a)$ is a good choice to get viable models\cite{Tamanini}. The torsion scalar $T=-6H^2$, where $H=\dot{a}/a$ is the Hubble parameter. Then the equations of motion (\ref{fieldequation}) give rise to the modified Friedmann equations
\begin{equation}\label{modifiedFriedmann}
f_T H^2=-\frac 4 3\pi G\epsilon-\frac 1 {12}f
\end{equation}
and
\begin{equation}
4\dot{H}f_T=\left[f_\mathcal{T}\left (3\frac{\partial p}{\partial \epsilon}-1\right )+16\pi G\right](\epsilon+p)-3H \dot{f}_T,
\label{modifiedFriedmann1}
\end{equation}
which are consequently different from those in previous references for $f(T, \mathcal{T})$ theory. It is easy to confirm the energy conservation (\ref{econserv}) from Eq. (\ref{modifiedFriedmann}) and Eq.(\ref{modifiedFriedmann1}).

\section{Cardassian universe from $f(T,\mathcal{T})$ theory}

\subsection{The action of Cardassian models from $f(T,\mathcal{T})$ theory}
In Ref. \cite{Zhai}, we studied the cosmology of gravity with the Lagrangian in the forms of $\mathcal{L}\propto -T+\alpha \sqrt{-T}+f(T,\mathcal{L}_m)$ and $\mathcal{L}\propto -T+\beta T^{-1}+f(T,\mathcal{L}_m)$. In the first form, the square root term is easy to prove as null, so $\alpha$ is actually a free parameter, and hence the correction of this term will not affect the local gravity tests. Similar to Ref. \cite{Zhai}, here we choose
\begin{equation}\label{fTTaction}
f(T, \mathcal{T})=-T+\alpha\sqrt{-T}+g(\mathcal{T}).
\end{equation}
For dust matter, the pressure is $p=0$. Then from (\ref{lm}) we have $\epsilon(\rho)=\rho$, and Eq. (\ref{modifiedFriedmann}) reduces to
\begin{equation}\label{Careqn}
H^2=\frac {8\pi G} 3 \rho+\frac 1 6g(\mathcal{T}),
\end{equation}
where $\mathcal{T}=-\rho$ for dust matter,
and $\rho\propto a^{-3}$ from Eq.\eqref{econserv}.
It is obvious that Eq. (\ref{Careqn}) is the very equation for Cardassian models and it is easy to find the forms of $f(T, \mathcal{T})$ corresponding to specific Cardassian models. Here, we examine three Cardassian models. The units $8\pi G=1$ is used hereinafter. For the original Cardassian model (OC)\cite{Freese},
\begin{equation}\label{OC}
H^2=\frac {\rho}3 \left[1+(\frac{\rho}{\rho_c})^{n-1}\right]
\end{equation}
where $\rho_c$ is the critical energy density at which the two terms of Eq.\eqref{OC} are equal, we have
\begin{equation}
	g(\mathcal{T})=2\rho(\frac{\rho}{\rho_c})^{n-1}=\frac2{\rho_c^{n-1}}\left( -\mathcal{T} \right)^n.
\end{equation}
For the modified polytropic Cardassian model (MPC)\cite{Freese3}
\begin{equation}\label{MPC}
H^2=\frac {\rho}3 \left[1+(\frac{\rho}{\rho_c})^{q(n-1)}\right]^{1/q},
\end{equation}
we have
\begin{equation}
	\begin{split}
g(\mathcal{T})=&2\rho\Big[\left[1+(\frac{\rho}{\rho_c})^{q(n-1)}\right]^{1/q}-1\Big]\\
=&2\mathcal{T}\Big[1-\left[1+(\frac{\mathcal{T}}{\rho_c})^{q(n-1)}\right]^{1/q}\Big],
\end{split}
\end{equation}
and for the exponential Cardassian model (EC)\cite{Liu}
\begin{equation}\label{EC}
H^2=\frac {\rho}3\exp\left[(\frac{\rho}{\rho_c})^{-n}\right],
\end{equation}
we have
\begin{equation}
	\begin{split}
g(\mathcal{T})=&2\rho\Big[\exp\left[(\frac{\rho}{\rho_c})^{-n}\right]-1\Big]\\
=&2\mathcal{T}\Big[1-\exp\left[(\frac{-\mathcal{T}}{\rho_c})^{-n}\right]\Big].
\end{split}
\end{equation}
Therefore, we claim that we find the possible origin of Cardassian models from $f(T,\mathcal{T})$ theory.

\subsection{$f(T,\mathcal{T})$-generalized Cardassian models}

Alternatively, inspired by the Lagrangian with the term $\beta T^{-1}$ considered in Ref. \cite{Zhai}, if we replace the $\alpha \sqrt{-T}$ term in Eq.(\ref{fTTaction}) with
\begin{equation}
-\frac{3\lambda^2H_0^4}{T},
\end{equation}
 we can obtain the $f(T,\mathcal{T})$-generalized Cardassian models.
For generalized OC (Model I), the modified FRW equation reads
\begin{equation}\label{equ:m1}
	E^2 -\frac{\lambda^2}{4}E^{-2} = \Omega_{0}(1+z)^3 + \Omega_x (1+z)^{3n}.
\end{equation}
Here $E(z) = \frac{H(z)}{H_0}$, $H_0$ is the Hubble parameter, $\Omega_0\equiv \frac{\rho_0}{3H_0^2} = \Omega_{m0} + \Omega_{b0}$, where $\Omega_{m0}$ and $\Omega_{b0}$
correspond to dark matter and baryons respectively, and
\begin{equation}\label{equ:ox}
	\Omega_x = 1- \frac{\lambda^2}{4}- \Omega_{0}.
\end{equation}

For the generalized MPC (Model II), the modified FRW equation reads
\begin{eqnarray}\label{equ:m2}
&	&E^2 -\frac{\lambda^2}{4}E^{-2}\nonumber\\
& =& \bigg\{\Omega_{0}^q(1+z)^{3q}  + \bigg[ \left(\Omega_x + \Omega_{0}\right)^q -\Omega_{0}^q \bigg] (1+z)^{3qn}\bigg\}^{1/q},
\end{eqnarray}
and for generalized EC (Model III), the modified FRW equation reads
\begin{eqnarray}\label{equ:m3}
&	&E^2 -\frac{\lambda^2}{4}E^{-2}\nonumber\\
& =& \Omega_{0}(1+z)^3 \exp{\bigg[ (1+z)^{-3n} \ln \left(\frac{\Omega_x + \Omega_{0} }{\Omega_{0}}\right)\bigg]}.
\end{eqnarray}

In all the cases, the modified FRW equations can be expressed unifiably as
\begin{equation}\label{equ:friedall}
	E^2 = \frac{1}{2} \bigg[ \phi(z) + \sqrt{\phi^2(z) + \lambda^2} \bigg]
\end{equation}
where $\phi(z)$ is the right hand side of Eqs.(\ref{equ:m1}), (\ref{equ:m2}), or (\ref{equ:m3}).

\subsection{ Observational Constraints}

In this subsection, using the observational data of SNeIa, CMB, and BAO, we give constraints and the best fit parameters of each model. For SNeIa data, we use the joint light-curve analysis(JLA)  sample, which contains $740$  spectroscopically confirmed type Ia supernovae with high quality light curves. The distance estimator in this analysis assumes that supernovae with identical colors, shapes, and galactic environments have, on average, the same intrinsic luminosity for all redshifts. This hypothesis is quantified by a linear model, yielding a standardized distance modulus\cite{Betoule:2014frx,Shafer:2015kda}
\begin{equation}\label{equ:modulusobs}
	\mu_{\text{obs}} = m_{\text{B}} - (M_{\text{B}} - A \cdot s + B \cdot C + P \cdot \Delta_M)  \,,
\end{equation}
where $m_{\text{B}}$ is the observed peak magnitude in rest-frame B band, and $M_{\text{B}}, s, C$ are the absolute magnitude, stretch, and color measures, which are specific to the light-curve fitter employed, and $P(M_* >10^{10} M_\odot)$ is the probability that the supernova occurred in a high-stellar-mass host galaxy. The stretch, color, and host-mass coefficients ($A, B, \Delta_M$, respectively) are nuisance parameters that should be constrained along with other cosmological parameters.

The CMB temperature power spectrum is sensitive to the matter density, and it also precisely measures the angular diameter distance $\theta_*$ at the last-scattering surface. We use the Planck measurement of the CMB temperature fluctuations and the WMAP measurement of the large-scale fluctuations of the CMB polarization. This CMB data are often denoted by "Planck + WP". The geometrical constraints inferred from this data set are the present values of baryon density $\Omega_{b0}h^2$, dark matter $\Omega_{m0}h^2$, and $\theta_*$ \cite{Zhai}, where $h$ is given by  $H_0= 100 h\:\text{km s}^{-1} \text{ Mpc}^{-1}$.

The BAO measurement provides a standard ruler to probe the angular diameter distance versus the redshift relation by performing a spherical average of their scale measurements,
see Ref. \cite{Eisenstein:1997ik}. We use the measurement of the BAO scale from Refs.~\cite{Beutler:2011hx,Padmanabhan:2012hf,Anderson:2012sa}.

In Table \ref{table:best}, we present the  best-fit parameters by using the data of CMB+BAO+JLA, and  also quote their $1\sigma$ bounds from the approximate Fisher information matrix. We also examine the constraints on parameters from the $1\sigma$ to the $3\sigma$ confidence levels for each model and Figs. 1-3 are the illustrations of the constraints on $\Omega_{m0}$ and $n$ for Models I, II, and III, respectively.

\begin{widetext}
\begin{table}[h]
\centering
  \begin{tabular}{c|c|c|c|c|c|c||c}
  \hline
  \hline
  \multirow{2}{*}{ Parameters} & \multicolumn{6}{c}{ Cosmological Models  } \\
  \cline{2-8}
  & OC &  Model I &  MPC & Model II &  EC &Model III & $\Lambda CDM$  \\
  \hline
  \hline
  $\Omega_{m0}$
  & $0.255^{+0.009}_{-0.010}$  	& $0.255^{+0.010}_{-0.010}$ & $0.256^{+0.011}_{-0.009}$ 	& $0.256^{+0.011}_{-0.011}$ &  $0.254^{+0.010}_{-0.010}$ & $0.251^{+0.010}_{-0.009}$ & $0.257^{+0.009}_{-0.009}$ \\
    $n$
  & $-0.022^{+0.052}_{-0.054}$	& $-0.014^{+0.062}_{-0.055}$ & $0.166^{+0.088}_{-0.098}$ 	& $0.377^{+0.102}_{-0.123}$ &  $0.720^{+0.039}_{-0.035}$ & $0.639^{+0.073}_{-0.079}$ &  $-$			     \\
  $q$
  & $-$                          	& $-$            		   & $1.387^{+0.257}_{-0.222}$ & $1.768^{+0.449}_{-0.397}$ &  $-$ 		     & $-$  			     & $-$			     \\
  $\lambda$
  & $-$ &$0.283^{+0.242}_{-0.246}$ 			   & $-$ &$0.915^{+0.278}_{-0.345}$ &  $-$ &$1.237^{+0.151}_{-0.208}$ & $-$\\
  \hline
   $H_0$
  & $68.46^{+1.232}_{-1.197}$ & $68.46^{+1.213}_{-1.239}$ & $68.55^{+1.227}_{-1.318}$ & $68.55^{+1.379}_{-1.285}$ &  $68.02^{+1.320}_{-1.241}$ & $68.77^{+1.299}_{-1.389}$ & $67.98^{+0.736}_{-0.737}$ \\
  $\Omega_{b0}h^2$
  & $0.0221\pm 0.0003$ & $0.0221\pm0.0003$ & $0.0220\pm 0.0003$ & $0.0220\pm 0.0003$ &  $0.0222\pm 0.0003$ & $0.0221\pm 0.0003$ & $0.0221\pm 0.0002$ \\
  \hline
  $A$
  & $0.141^{+0.007}_{-0.006}$ & $0.141^{+0.007}_{-0.006}$ & $0.140^{+0.006}_{-0.007}$ & $0.141^{+0.007}_{-0.007}$ &  $0.142^{+0.006}_{-0.007}$ & $0.142^{0.007}_{-0.006}$ &$0.141^{+0.007}_{-0.006}$ \\
  $B$
  & $3.103^{+0.083}_{-0.079}$ & $3.103^{+0.088}_{-0.085}$ & $3.101^{+0.082}_{-0.087}$ & $3.101^{+0.078}_{-0.085}$ &  $3.112^{+0.079}_{-0.079}$ & $3.112^{+0.086}_{-0.083}$ &$3.100^{+0.082}_{-0.086}$ \\
  $M_B$
  & $-19.10^{+0.031}_{-0.031}$ & $-19.10^{+0.031}_{-0.032}$ & $-19.09^{+0.032}_{-0.032}$ & $-19.09^{+0.038}_{-0.035}$ &  $-19.14^{+0.031}_{-0.033}$ & $-19.11^{+0.035}_{-0.036}$ &$-19.11^{+0.026}_{-0.026}$ \\
  $\Delta M$
  & $-0.070^{+0.022}_{-0.025}$ & $-0.070^{+0.022}_{-0.024}$ & $-0.070^{+0.022}_{-0.021}$ & $-0.070^{+0.023}_{-0.022}$ &  $-0.069^{+0.022}_{-0.022}$ & $-0.069^{+0.023}_{-0.021}$ &$-0.070^{+0.023}_{0.023}$ \\
  \hline
  \hline
  $\chi^2_{min}/d.o.f$
  & $683.908/738$ &$683.907/737$ & $683.616/737$ &$683.590/736$ &$688.767/738$ & $685.693/737$ & $684.131/739$\\
  \hline
  \hline
  \end{tabular}
  \caption{\label{table:best} Best fitting parameters for all the models.}
\end{table}
\end{widetext}

\begin{figure}[h]
\begin{center}
\includegraphics[width=0.4\textwidth,angle=0]{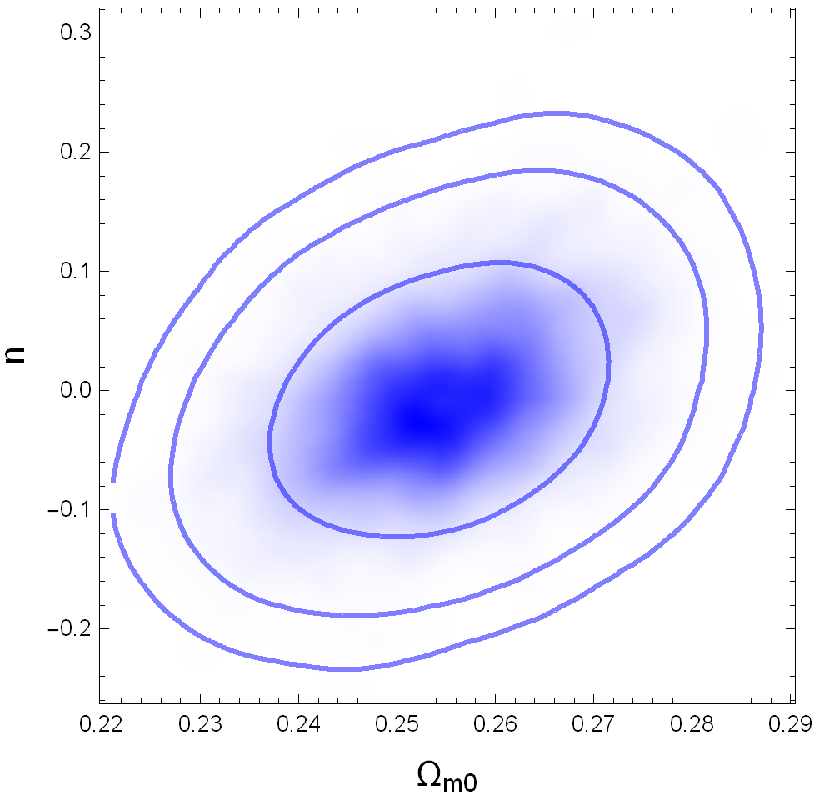}
\caption{Constraints on $\Omega_{m0}$ and $n$ from the $1\sigma$ to the $3\sigma$ confidence levels by using JLA SNe Ia + CMB + BAO for  model I, while other parameters take their best fitting values. }
\end{center}
\end{figure}

\begin{figure}[h]
\begin{center}
\includegraphics[width=0.4\textwidth,angle=0]{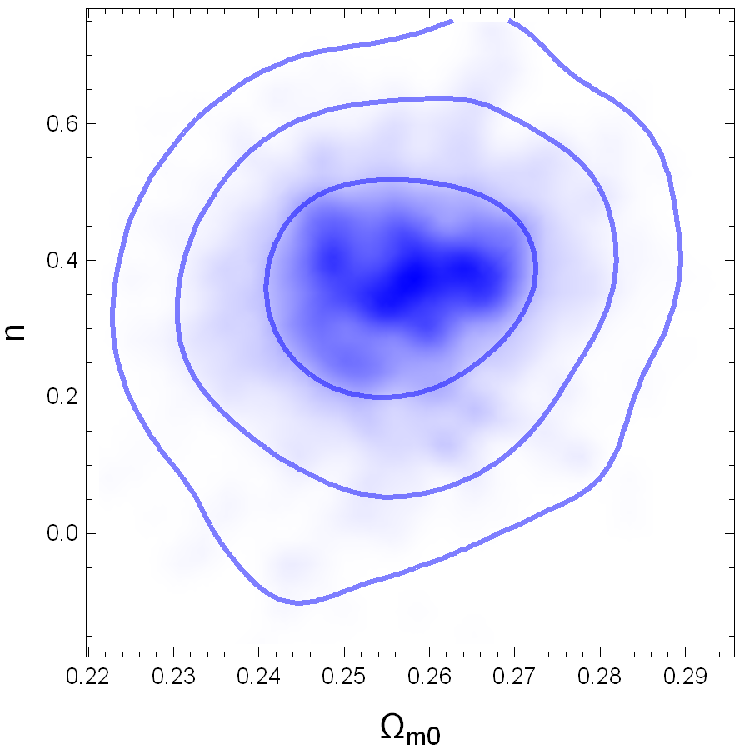}
\caption{Constraints on $\Omega_{m0}$ and $n$ from the $1\sigma$ to the $3\sigma$ confidence levels by using JLA SNe Ia + CMB + BAO for  model II, while other parameters take their best fitting values. }
\end{center}
\end{figure}

\begin{figure}[h]
\begin{center}
\includegraphics[width=0.4\textwidth,angle=0]{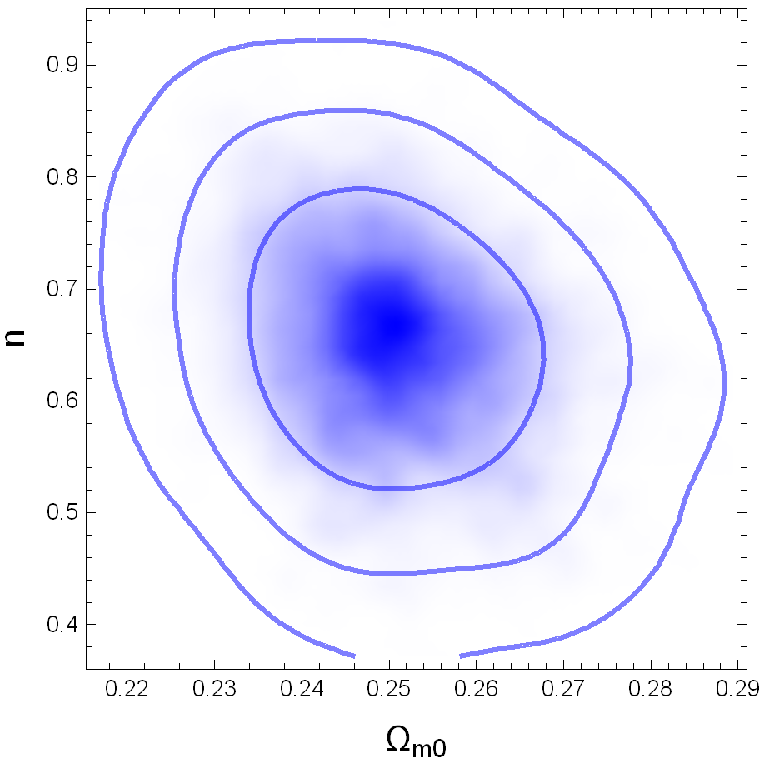}
\caption{Constraints on $\Omega_{m0}$ and $n$ from $1\sigma$ to $3\sigma$ confidence level by using JLA SNe Ia + CMB + BAO for  model III, while other parameters take their best fitting values. }
\end{center}
\end{figure}

 \section{conclusion and discussion}

Using the result of the Lagrangian of a barotropic fluid given in Ref.\cite{Minazzoli}, we rederive $f(T,\mathcal{T})$ gravity, obtaining the modified Friedmann equations. We find the connection between $f(T,\mathcal{T})$ gravity and the Cardassin universe. For dust matter, the modified Friedmann equations from $f(T,\mathcal{T})$ theory can correspond to those of the Cardassian models, and thus, a possible origin of the Cardassian universe is given. We present the Lagrangians of the original Cardassian model, the modified polytropic Cardassian model, and the exponential Cardassian model from $f(T,\mathcal{T})$ theory. Furthermore, we get generalized Cardassian models by adding an additional term to the corresponding Lagrangians of $f(T,\mathcal{T})$ theory that lead to the three Cardassian models mentioned above. Using the data of CMB+BAO+JLA, we get the fitting results of the cosmological parameters and give constraints of model parameters for all these models.

As one of the candidates for explaining the acceleration of the universe, Cardassian models have advantages in that the universe can be flat and yet consist of only matter and radiation, both  of which satisfy the conservation laws. However, there is not a satisfactory answer in the literature for the origin of the Cardassian models. In our new derivation of $f(T,\mathcal{T})$ theory, the usual energy conservation still holds, which is necessary for Cardassian models. The conclusion that we have given a possible origin of the Cardassian universe from $f(T,\mathcal{T})$ gravity is thus consistent. The connection we have found between the two theories is interesting and will be good in seeking the explanation of the accelerated expansion of the universe.

\begin{acknowledgments}
This work is supported by the National Science Foundation of China, Grants No.~10671128, No.~11105091 and No.~11047138, and the Key Project of Chinese Ministry of Education, Grant No.~211059.
\end{acknowledgments}

\end{document}